\documentclass[aps,superscriptaddress,twocolumn,twoside,floatfix,pra,nofootinbib,a4paper]{revtex4-2}
\usepackage{times}
\usepackage{amsmath,amssymb,amsthm,mathtools,thmtools,nameref}
\usepackage{comment}
\usepackage{enumerate}
\usepackage{float}
\usepackage[dvipsnames]{xcolor}
\usepackage[colorlinks=true,linkcolor=blue,citecolor=magenta,urlcolor=blue]{hyperref}
\usepackage[framemethod=default]{mdframed}
\usepackage{csvsimple}
\usepackage[mode=tex]{standalone}
\usepackage[capitalize]{cleveref}
\usepackage{multirow}
\usepackage{bigdelim}
\usepackage{booktabs}
\usepackage{quantikz}
\usepackage{colortbl}
\usepackage{nicefrac}
\usepackage[table]{xcolor}

\usepackage{graphicx}	
\usepackage{subcaption}
\usepackage[margin=3cm]{geometry}
\usepackage{url}
\usepackage{todonotes}
\usepackage{bbm}

\usepackage{tikz}

\usepackage{pifont}
\usepackage{makecell}
\usepackage{adjustbox}

\usepackage{epsfig}
\usetikzlibrary{shapes.symbols,patterns} % for source symbols
\usepackage{pgfplots}
\pgfplotsset{compat=1.10}
\usepgfplotslibrary{fillbetween}

\newcommand*{\eps}{\varepsilon}
\newcommand*{\cE}{\mathcal{E}}

\definecolor{turquoise}{HTML}{00E7C0}

\begin{document}

\title{Practical Entanglement Distillation Protocols with Quadratic Error Suppression}
\author{Elisa~B\"aumer~Marty}
\email{eba@zurich.ibm.com}
\affiliation{\small IBM Quantum, IBM Research Europe -- Zurich}
%\date{\today}

\begin{abstract}
Near-term and early fault-tolerant quantum computing architectures are expected to exhibit highly non-uniform error rates. In particular, local operations within a chip can be substantially more reliable than operations connecting different chips or dilution refrigerators. Such inter-module operations can therefore become a dominant bottleneck, even when quantum error correction is applied.
Entanglement distillation provides a natural way to trade additional operations and qubits for higher-fidelity entanglement. Standard distillation protocols, however, are usually formulated in an LOCC resource model, in which several noisy Bell pairs are generated initially and all subsequent processing consists only of local operations and classical communication. Here, we consider a generalized model tailored to modular quantum computing hardware, in which the two modules have access to high-fidelity local operations and to repeated uses of the same noisy inter-module entangling operation during the protocol.
We develop practical small-scale entanglement distillation protocols designed to minimize both space and time overhead. Remarkably, our main protocol requires only two qubits per module, yet achieves quadratic error suppression of inter-module errors, assuming local operations are much cleaner.
Compared with existing small-scale protocols, our space-optimal protocol provides more space- and time-efficient quadratic error suppression and achieves the best performance in our simulations and experiments on noisy links of current superconducting quantum processors. These results suggest that inter-module-gate-assisted entanglement distillation can be a practical primitive for overcoming noisy links in modular quantum computing architectures.
\end{abstract}

\maketitle

\section{Introduction}
Over the past decade, substantial progress has been made toward scalable quantum computers, with increasing qubit numbers and physical error rates reduced by orders of magnitude. These advances have enabled demonstrations of quantum utility~\cite{kim2023utility} and motivated quantum-centric supercomputing architectures~\cite{mezzacapo2025}. 
However, error rates can vary substantially across different operations due to differences in coherence times, readout errors and gate fidelities. It becomes especially important in modular architectures with a limited number of qubits per dilution refrigerator, where operations connecting different chips or fridges are required. Such inter-module operations are expected to be significantly noisier than local operations, as is typical in modular or networked quantum computing proposals~\cite{Nickerson2014}.
Similar heterogeneity can also arise at the logical level. For example, QEC schemes that connect smaller code blocks may require inter-module measurements whose error rates are orders of magnitude larger than those of in-module measurements, as in proposals based on the Gross code~\cite{Yoder25tourdegross}. More generally, connecting error-corrected qubits through noisy links is a central challenge for fault-tolerant interconnect schemes~\cite{Ramette2024}.
\textit{Entanglement distillation} provides a natural way to address this bottleneck. The concept was introduced already 30 years ago~\cite{Bennett96} as a method for converting multiple noisy Bell pairs into fewer ones of higher fidelity. Entanglement distillation is closely related to quantum error correction~\cite{BennettQECC96} and is a key component of quantum repeater architectures for long-distance quantum communication~\cite{Briegel1998,Dur1999}.
However, the original protocol was primarily a proof-of-principle and thus not very efficient and more recent work has mostly focused on large-scale protocols that aim at optimizing asymptotic behavior~\cite{BennettQECC96,DevetakWinter2005,Pattison2024,Shi2025}, rather than practically-relevant small-scale protocols. In particular, most protocols would perform poorly in regimes where input error rates are already small, or require large space or time overheads, especially since the resources that were optimized were typically not space and time, but the number of input Bell pairs. These objectives are different from the requirements of near-term and early fault-tolerant quantum computing hardware, where the relevant resources are typically qubit count, circuit depth and wall-clock time.

Here, we use a generalized resource model motivated by modular quantum computing hardware. Standard entanglement distillation protocols are usually formulated as LOCC protocols, in which several noisy Bell pairs are first distributed between two parties and the subsequent processing uses only local quantum operations, measurements and classical communication. In modular quantum processors, however, the same operation that creates a raw Bell pair may also be available as a noisy inter-module entangling gate during the protocol. We therefore allow repeated uses of this noisy inter-module operation, in addition to high-fidelity local operations, measurements and classical communication. While this setting differs from standard LOCC distillation, it captures the relevant cost tradeoff for architectures in which inter-module operations are expensive and noisy, while local operations are comparatively cheap and reliable.

Within this setting, we introduce practical entanglement distillation protocols that minimize qubit overhead and are among the fastest protocols with quadratic suppression of inter-module errors, meaning that the leading-order contribution of the output error scales as $\mathcal{O}(\eps_{inter}^2)$. Our main protocol requires as little as two qubits per module, which makes it space-optimal in the sense that no protocol producing a purified Bell pair while detecting inter-module faults can use fewer than two qubits per module: one qubit per module is needed to hold the output Bell pair, and an additional qubit per module is needed to extract error information.
We refer to the noisier operations, such as noisy links within a chip, or long-range gates connecting multiple chips, as inter-module operations, and to the higher-fidelity operations as local operations. When inter-module error rates are orders of magnitude larger than local error rates, the proposed protocols can reduce the corresponding bottleneck and thus be advantageous for both near-term devices and fault-tolerant architectures. The key idea of our improved protocols is that inter-module operations can be used as active components for the distillation circuit, rather than only for preparing initial noisy Bell pairs.

We start by introducing the noise model in~\cref{sec:noisemodel}, followed by an overview of existing protocols in~\cref{sec:old_protocols}. We then present our new protocols in~\cref{sec:new_protocols} and compare their error suppression rates, qubit numbers, and duration in~\cref{sec:comparison}. In~\cref{sec:demonstrations}, we demonstrate their efficiency both with simulations for varying input errors as well as with experiments on noisy links on current devices. Finally, we conclude in~\cref{sec:conclusion}.

\section{Distillation Protocols and Analytical Comparison}

\subsection{Noise model} \label{sec:noisemodel}
Arbitrary noise during an entangling gate can be described by a general CPTP map: $\cE(\rho)=\sum_kE_k\rho E_k^\dag$, where $E_k$ are Kraus operators acting on two qubits. By applying Pauli twirling through randomized compiling~\cite{Wallman2016RandomizedCompiling}, the ensemble-averaged effective noise channel is transformed into a stochastic Pauli channel~\cite{Dankert2009}, such that in the Pauli-basis representation the off-diagonal terms associated with coherent noise are averaged away, leaving only diagonal Pauli-error terms:
\begin{align}
    \tilde{\cE}(\rho)=\sum_{P\in\{I,X,Y,Z\}^{\otimes 2}} \eps_P P \rho P \,,
\end{align}
with $\sum_P \eps_P = 1$. 
This has 15 independent parameters, but for compact analytic expressions, we also consider a two-parameter weight-based Pauli error model in which all six single-qubit errors occur with probability $\frac{\eps_1}{6}$, and all nine two-qubit correlated Pauli errors occur with probability $\frac{\eps_2}{9}$. This simplification does not capture the full structure of general noise, but it can still be used to obtain conservative upper bounds by assigning all single-qubit (two-qubit) Pauli errors the maximal probability within their class. Denoting the maximum probability of any single-qubit (two-qubit) Pauli error by $p^{(1)}_{\max}$ ($p^{(2)}_{\max}$), this yields the bounds $\eps_1\geq 6 p^{(1)}_{\max}$ and $\eps_2\geq 9 p^{(2)}_{\max}$.
However, these bounds are typically not tight, as realistic noise can be highly non-uniform and protocol performance can depend on the specific distribution of Pauli errors.

When a Bell state is created via a noisy entangling gate, any correlated errors on the two qubits effectively result in a single-qubit error on either qubit.  
Thus, we can describe any noise on a Bell state as a classical mixture of $X$, $Z$ and $Y$ errors on one of the qubits, where $Y$ is equivalent to the combined action of $X$ and $Z$. Taking the uniform weight-based Pauli error model from above, we can write the noisy Bell state in the basis of the four Bell states $\ket{\Phi^+}=\frac{1}{\sqrt{2}}(\ket{00}+\ket{11})$, \mbox{$\ket{\Phi^-}=\frac{1}{\sqrt{2}}(\ket{00}-\ket{11})$}, $\ket{\Psi^+}=\frac{1}{\sqrt{2}}(\ket{01}+\ket{10})$ and $\ket{\Psi^-}=\frac{1}{\sqrt{2}}(\ket{01}-\ket{10})$ as
\begin{align} \label{eq:noisyBell}
    \rho_B = (1-\eps_B) &\ket{\Phi^+}\bra{\Phi^+} + \frac{\eps_B}{3}\ket{\Phi^-}\bra{\Phi^-} \nonumber\\
    + \frac{\eps_B}{3} &\ket{\Psi^+}\bra{\Psi^+} + \frac{\eps_B}{3}\ket{\Psi^-}\bra{\Psi^-}\,,
\end{align}
with a total error given by 
\begin{align}
    \eps_{B} = \eps_1+\frac{2}{3} \eps_2 \,.
\end{align}
For protocols with a single output Bell pair, the output error is defined as the probability that the output state differs from $\ket{\Phi^+}$. For protocols producing multiple Bell pairs, we define the output error as the joint error, i.e. the probability that at least one of the output Bell pairs is incorrect.

Entanglement distillation protocols are only beneficial when some operations have a much smaller error than others. We refer to those in the following as \textit{local operations} and to the operations that we assume to have an error orders of magnitude larger as \textit{inter-module operations}, as they are likely to appear on links between chips or fridges, even though they may also just be noisy links within a chip.
In the circuit diagrams in~\cref{fig:recurrence,fig:spaceopt} we denote the potential inter-module errors with different colors  and indicate how they propagate to the end of the circuit, where a single error can be any combination of 'X' and 'Z' of the same color. 
As we assume that the local gates and measurements have an error that is orders of magnitude smaller than that of the inter-module gates, we take them to be perfect in our analysis in this section. In practice the local errors will yield a lower bound on the distillable output error rate as we will see in the simulations and experiments in~\cref{sec:demonstrations}.
Throughout the analytical comparison, all output error rates are reported to leading non-vanishing order in the inter-module error probabilities, conditioned on successful postselection.
Higher-order contributions are omitted. Thus, for protocols with linear error suppression we keep terms of order $\mathcal{O}(\eps)$, whereas for protocols with quadratic error suppression we report the leading $\mathcal{O}(\eps^2)$ contribution.

\subsection{Existing protocols}\label{sec:old_protocols}
\begin{figure*}[!htb]
\centering
\includegraphics[width=2.\columnwidth]{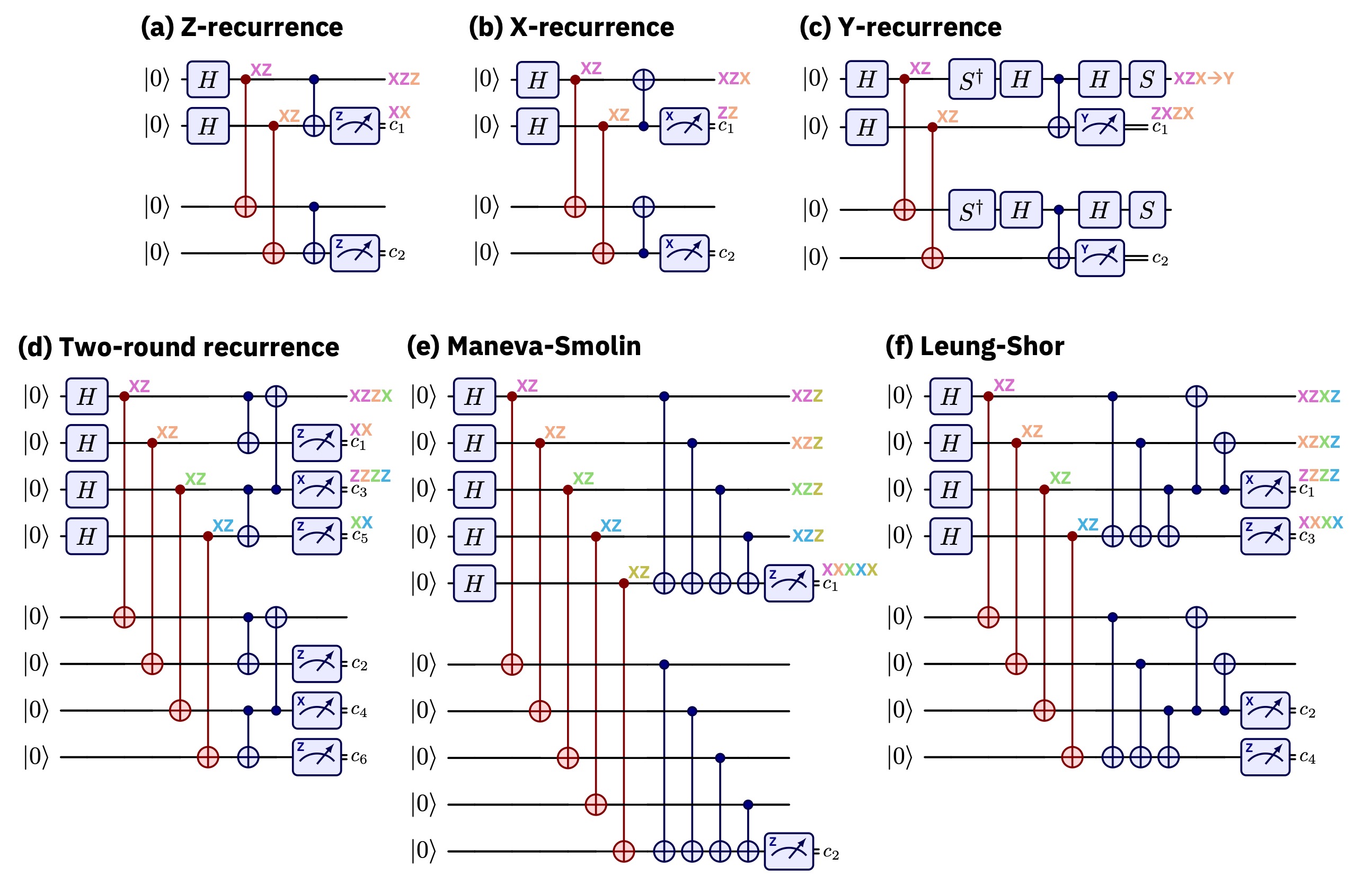}
\caption{Existing protocols. Blue operations are local, red operations are noisy inter-module operations. The protocol accepts if $c_1=c_2$, $c_3=c_4$ and $c_5=c_6$. \textbf{(a)-(c)} Single recurrence checks in the $Z$, $X$ and $Y$ bases, each detecting two of the three Pauli error types.  \textbf{(d)} Two recurrence rounds (here Z and X) detect any single error. \textbf{(e)} Maneva-Smolin protocol, which increases the output rate by checking multiple Bell pairs jointly. \textbf{(f)} Leung-Shor protocol, which outputs two Bell pairs with quadratic error suppression.} 
\label{fig:recurrence}
\end{figure*}
In the following, the quoted error rates are acceptance-conditioned and given to leading non-vanishing order in the input error probabilities, following the convention introduced in~\cref{sec:noisemodel}.

\paragraph{Recurrence protocols}
The original entanglement distillation protocol introduced in~\cite{Bennett96}, often called \textit{recurrence}, is illustrated in~\cref{fig:recurrence}(a)-(c). It checks the parity of two Bell pairs, either in the $X$, $Z$ or $Y$ basis. While it can be used to detect errors of a specific form, it cannot detect an arbitrary single-qubit error in one round. For example, the Z-recurrence protocol detects a single bit flip, but not a phase flip. Taking two noisy Bell pairs as input, each of the form~\cref{eq:noisyBell} with error $\eps_B$, the output after one round is one noisy Bell state with error $\frac{2}{3}\eps_B$.

However, as described in~\cite{Deutsch96}, one can combine two rounds of recurrence to detect any single error, as illustrated in~\cref{fig:recurrence}(d) (note that the original protocol used a different change of basis, but the idea and final error rates are the same). This requires four noisy Bell pairs and returns one Bell pair with quadratically suppressed error $\frac{8}{9}\eps_B^2$. Trading time for space, one could also apply the Z-checks sequentially and reduce the number of qubits from four to three on each block. There have been efforts optimizing the sequence of such recurrence protocols for different noise~\cite{Abdelhadi26}.

\paragraph{Maneva-Smolin protocol}
To increase the number of output Bell pairs per round, the check can be performed on multiple Bell pairs together, as proposed in~\cite{Maneva2000} and illustrated in~\cref{fig:recurrence}(e) (here shown for $m=5$ Bell pairs, but it works for any $m\geq2$). Using $m$ noisy Bell pairs, this protocol outputs $m-1$ higher-fidelity Bell pairs. The Maneva-Smolin protocol was designed to improve the yield in the limit of arbitrarily large numbers of states, but like a single recurrence round, it only detects certain types of errors (here X and Y errors). For Bell states of the form~\cref{eq:noisyBell} with error $\eps_B$, the marginal error on each output Bell state is thus $\frac{2}{3}\eps_B$. However, the errors are correlated across output Bell pairs. The joint error, i.e. the probability that at least one of the $m-1$ outputs is incorrect, scales as $\frac{m}{3}\eps_B$. A straightforward second round that circumvents the correlated noise would then use $m$ batches of $m$ Bell pairs each, i.e. would in total require $m^2$ noisy Bell pairs to obtain $(m-1)^2$ higher-fidelity Bell pairs with quadratic error suppression.

\paragraph{Leung-Shor protocol}
In~\cite{Leung2007} an improved protocol for entanglement distillation was presented that can detect any single error. As shown in~\cref{fig:recurrence}(f), it takes four noisy Bell states as input and outputs two higher-fidelity Bell states with quadratic error suppression. It can be seen as starting with the Maneva-Smolin protocol~\cite{Maneva2000} for $m=4$ in the Z-basis and then applying it again on the three output qubits in the X-basis. 
Crucially, for four input Bell pairs the error correlations generated by the first step do not leave any undetected single-error contributions after the second step, unlike what would occur for, e.g., $m=5$. The protocol outputs two Bell pairs with correlated errors. The joint error, i.e. the probability that at least one of the outputs is incorrect, is $\frac{14}{9}\eps_B^2$.

\subsection{Our more efficient protocols}\label{sec:new_protocols}

\paragraph{Space-optimal protocols} \label{sec:spaceopt}
\begin{figure*}[!htb]
\centering
\includegraphics[width=2.\columnwidth]{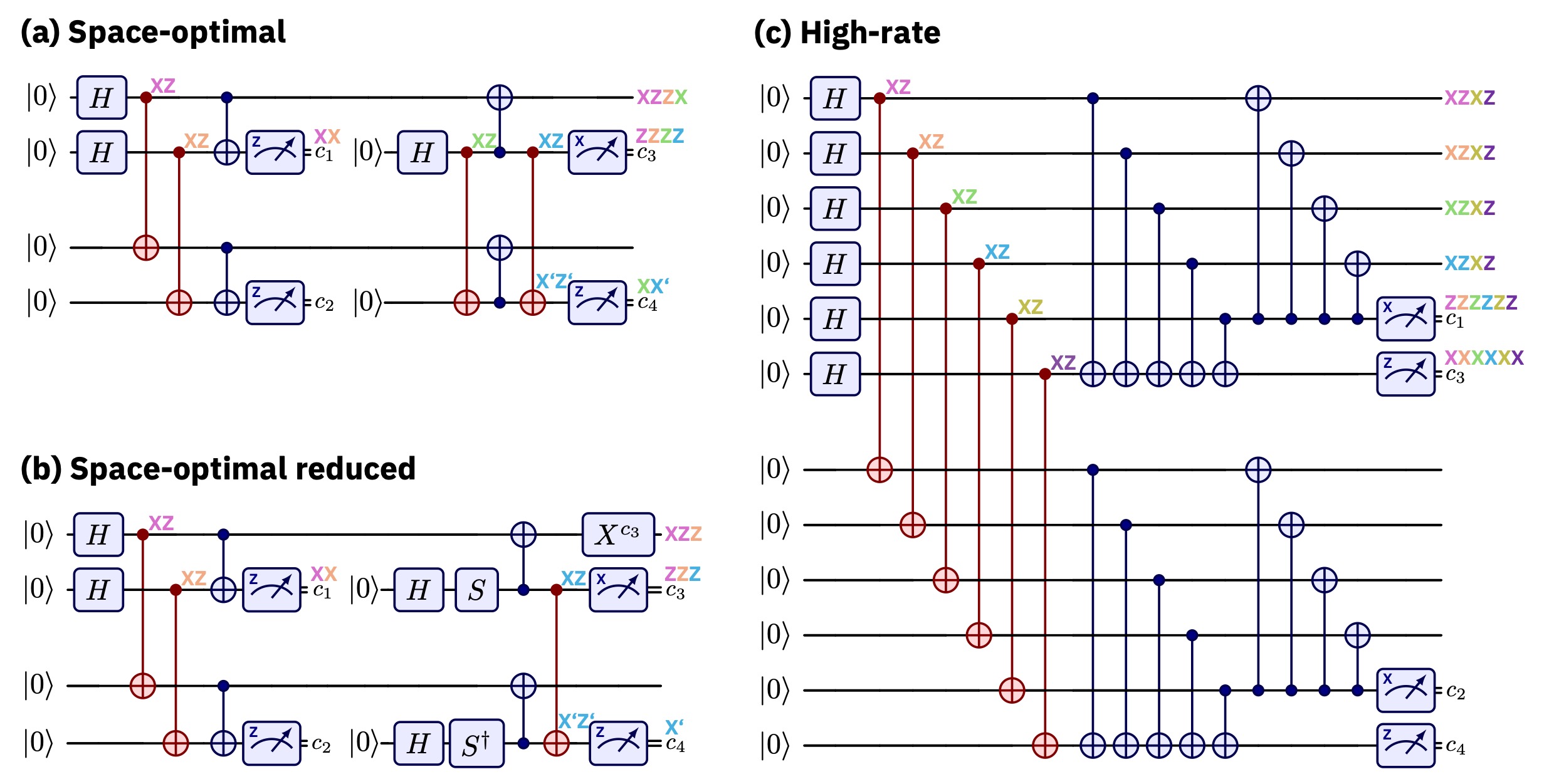}
\caption{Our new efficient protocols. Blue operations are local, red operations are noisy inter-module operations. The protocol accepts if $c_1 = c_2$ and $c_3=c_4$, and in \textbf{(a)} additionally requires $c_3=c_4=0$.
\textbf{(a)} Space-optimal protocol with quadratic suppression of any two-qubit Pauli errors. \textbf{(b)} Reduced space-optimal protocol with fewer gates, achieving quadratic suppression only for arbitrary single-qubit Pauli errors. \textbf{(c)} High-rate protocol acting on $m$ qubits per module and outputting $m-2$ Bell pairs with quadratic error suppression.}
\label{fig:spaceopt}
\end{figure*}
We first introduce our main result: a space-optimal entanglement distillation protocol that requires only two qubits per module and outputs one Bell pair with quadratically suppressed error. The important difference is that unlike original protocols that would consume inter-module entanglement only via Bell pairs, here we also apply an inter-module operation directly. One such inter-module operation is counted with the same cost used to create a Bell pair, but it enables more space- and time-efficient protocols. 

In ~\cref{fig:spaceopt}(a) we present our main space-optimal entanglement distillation protocol which requires four inter-module connections and can detect arbitrary single- and two-qubit Pauli errors. If the dominant inter-module faults are single-qubit errors, one inter-module gate could be removed, yielding the slightly faster reduced protocol shown in~\cref{fig:spaceopt}(b). Note that in addition to arbitrary single-qubit Pauli errors, it could also detect any ZZ- and XX-errors, but it fails to detect errors from the final inter-module gate that correspond to correlated Pauli errors of the form $\{Z,Y\}$ on the top qubit and $\{X,Y\}$ on the bottom qubit. Whether this is an error we should consider depends on the physical implementation of the inter-module gate. 

If one spare clean ancilla qubit on each side is available, both protocols could be further parallelized trading time for space. This can be especially useful for the case where measurements take significantly longer than two-qubit operations, as no measurement of qubits is required during the protocol anymore, but only at the end.

\paragraph{High-rate protocol} \label{sec:highrate}
A natural generalization combines the Maneva-Smolin and the Leung-Shor constructions by extending the protocol to larger $m$, as illustrated for $m=6$ in~\cref{fig:spaceopt}(c). However, due to the correlated errors, the detection of an arbitrary error only works for an even number of input Bell states $m$ and then outputs $m-2$ Bell pairs with quadratically suppressed error $(\frac{1}{6}m^2-\frac{7}{18}m+\frac{4}{9})\eps_B^2$. This requires at least four Bell pairs (in which case it corresponds to the Leung-Shor protocol), but yields a higher output rate for increasing number of qubits. While in most applications we expect the number of qubits per module to be very limited, this protocol can find an application whenever multiple clean qubits per module are available, in particular in the very beginning of an algorithm.

\subsection{Comparison} \label{sec:comparison}
\begin{table*}[htb]
    \centering
    \renewcommand{\arraystretch}{1.5} 
    \begin{adjustbox}{width=1.\textwidth,center}
    \begin{tabular}{l|ccccccc}
\makecell[c]{\textbf{Protocol}}
& \makecell[c]{\textbf{\# qubits}\\\textbf{per}\\\textbf{module}}
& \makecell[c]{\textbf{Depth}\\\textbf{1-D}}
& \makecell[c]{\textbf{Depth}\\\textbf{all-to-all}}
& \makecell[c]{\textbf{Duration}\\{}\textbf{[ns]}}
& \makecell[c]{\textbf{\# output}\\\textbf{Bell pairs}}
& \makecell[c]{\textbf{\# inter-}\\\textbf{module}\\\textbf{operations}}
& \makecell[c]{\textbf{Output}\\\textbf{error rate}} \\\toprule
    \raisebox{1pt}{Single recurrence} & \multirow{2}{*}[3pt]{$2$} & \multirow{2}{*}[3pt]{$5~(+1)$} & \multirow{2}{*}[3pt]{$2~(+1)$} & \multirow{2}{*}[3pt]{$928~(+1320)$} & \multirow{2}{*}[3pt]{$1$} & \multirow{2}{*}[3pt]{$2$} & \multirow{2}{*}[3pt]{\textcolor{red}{$\frac{2}{3}(\eps_1 + \frac{2}{3}\eps_2)$}}\\ [-5pt] 
    \raisebox{3pt}{~[\cref{fig:recurrence}(a)]~\cite{Bennett96}} \\ [-3pt]\hline
    \raisebox{-4pt}{Two-round recurrence} & $3$ & $10~(+2)$ & $5~(+2)$ & $3296~(+1288)$ & \multirow{2}{*}{$1$} & \multirow{2}{*}{$4$} & \multirow{2}{*}{$ \frac{8}{9}(\eps_1+\frac{2}{3}\eps_2)^2$} \\ 
    \raisebox{4pt}{~[\cref{fig:recurrence}(d)]~\cite{Deutsch96}} & $4$ & $14~(+1)$ & $3~(+1)$ & $2688~(+1288)$ & & & \\ \hline
    \raisebox{1pt}{Maneva-Smolin} & \multirow{2}{*}[3pt]{$m$} & \multirow{2}{*}[3pt]{$10m-18$} & \multirow{2}{*}[3pt]{$m~(+1)$} & \multirow{2}{*}[3pt]{see~\cref{fig:expresults}(c)} & \multirow{2}{*}[3pt]{$m-1$} & \multirow{2}{*}[3pt]{$m$} & \multirow{2}{*}[3pt]{\textcolor{red}{$ \frac{m}{3}(\eps_1+\frac{2}{3}\eps_2)$}} \\  [-5pt]
    \raisebox{3pt}{~[\cref{fig:recurrence}(e)]~\cite{Maneva2000}} & \\ [-3pt] \hline
    \raisebox{1pt}{Leung-Shor} & \multirow{2}{*}[3pt]{$4$} & \multirow{2}{*}[3pt]{$19~(+2)$} & \multirow{2}{*}[3pt]{$4~(+2)$} & \multirow{2}{*}[3pt]{$3240~(+1440)$} & \multirow{2}{*}[3pt]{$2$} & \multirow{2}{*}[3pt]{$4$} & \multirow{2}{*}[3pt]{$\frac{14}{9}(\eps_1+\frac{2}{3}\eps_2)^2$} \\[-5pt] 
    \raisebox{3pt}{~[\cref{fig:recurrence}(f)]~\cite{Leung2007}} \\ [-3pt]\hline 
    \rowcolor{turquoise!8}[1\tabcolsep]\raisebox{-4pt}{Space-optimal} & $\textbf{2}$ & $8~(+2)$ & $5~(+2)$ & $2608~(+1376)$ & & & \\ 
    \rowcolor{turquoise!8}[1\tabcolsep]\raisebox{4pt}{~[\cref{fig:spaceopt}(a)]~[ours]} & $3$ & $11~(+1)$ & $3~(+2)$ & $1856~(+1320)$ & \multirow{-2}{*}{1} & \multirow{-2}{*}{$4$} & \multirow{-2}{*}{$\frac{7}{9}\eps_1^2+\frac{4}{9}\eps_2^2+\frac{32}{27}\eps_1 \eps_2$} \\ \hline
    \raisebox{-4pt}{Space-optimal reduced} & $2$ & $7~(+2)$ & $4~(+2)$ & $2520~(+1376)$ & \multirow{2}{*}{$1$} & \multirow{2}{*}{$3$} & \multirow{2}{*}{$\frac{2}{3}\eps_1^2+\textcolor{red}{\frac{4}{9}\eps_2}$}\\ 
    \raisebox{4pt}{~[\cref{fig:spaceopt}(b)]~[ours]} & $3$ & $6~(+4)$ & $3~(+2)$ & $688~(+1648)$ & & & \\ \hline    
    \raisebox{1pt}{High-rate} & \multirow{2}{*}[3pt]{$m$} & \multirow{2}{*}[3pt]{$8m\!-\!13~(+2)$} & \multirow{2}{*}[3pt]{$m~(+2)$} & \multirow{2}{*}[3pt]{see~\cref{fig:expresults}(c)} & \multirow{2}{*}[3pt]{$m-2$} & \multirow{2}{*}[3pt]{$m$} & \multirow{2}{*}[3pt]{$(\frac{1}{6}m^2\!-\!\frac{7}{18}m\!+\!\frac{4}{9}) (\eps_1\!+\!\frac{2}{3}\eps_2)^2$} \\ [-5pt]   
    \raisebox{3pt}{~[\cref{fig:spaceopt}(c)]~[ours]} & & \\ [-3pt] \hline
    \end{tabular}
    \end{adjustbox}
    \caption{Comparison of the existing and new protocols. The shaded row shows our main result. Output error rates are conditioned on successful postselection and reported to leading order in the inter-module error probabilities. For multi-output protocols, the reported error is the joint error. Linear error suppression is colored red due to its inefficiency. The individual columns are described in detail in~\cref{sec:comparison}}
    \label{tab:comparison}
\end{table*}

To highlight the advantage of our new entanglement distillation protocols, we compare them in detail in~\cref{tab:comparison}.
For most protocols, we consider both the more space-efficient sequential implementation and the more time-efficient parallel implementation and compare the following characteristics:
\begin{itemize}
    \item \textit{\# qubits per module}: The number of qubits that we require on each of the two modules we are connecting.
    \item \textit{Depth 1-D}: The number of time steps required when the protocol is compiled to a line, with the output Bell pair(s) ending on the outermost qubits. While this depends on the native gate set and is highly hardware-dependent, here we assume that both, two-qubit operations and measurements, require one time step and neglect single-qubit Clifford operations. The one-dimensional connectivity yields an upper bound on the depth.
    We also assume that we can use the Bell pair as soon as the last gate acted on it, as the postselection probability is relatively high and if a later check fails, any subsequent operations performed before receiving the postselection outcome can be reversed. In brackets we add the extra time required on the ancilla qubits. Note that to account for the postselection rate, one would need to divide by it, but since it is very close to one it becomes negligible.
    \item \textit{Depth all-to-all}: The number of time steps of the protocol given an all-to-all connectivity and allowing parallelization of all operations acting on different qubits. Analogously to the depth on a line, we assume that both, two-qubit operations and measurements, require one time step, neglect single-qubit Clifford operations and determine only the time until we can use the Bell pair, with brackets indicating the extra time required on the ancilla qubits. Allowing all-to-all connectivity and full parallelization yields a lower bound on the depth.
    \item \textit{Duration}: This is the duration of the protocol when compiled on a line on \texttt{ibm\_pittsburgh}. Here we account for all operations, but \mbox{analogously} to the depth determine only the time until we can use the Bell pair, with brackets indicating the extra time required on the ancilla qubits. Note that due to the fast gate times of $32$~ns for single- and $88$~ns for two-qubit gates compared to the relatively slow readout times of $1288$~ns, the orders according to the protocols' depths and durations differ. 
    \item \textit{\# output Bell pairs}: The number of output Bell pairs we get from each execution of the protocol.
    \item \textit{\# inter-module operations}: The number of inter-module gates required for each execution of the protocol. Using $k$ long-range gates with error $\eps$ lower bounds the postselection rate by $(1-\eps)^k$. We expect that they have a duration comparable to in-module operations and since the resulting reduction in the acceptance probability is small in the low-error regime, this has only a minor effect on the overall cost.
    \item \textit{Output error rate}: The final error rate after one successful round with noisy inter-module gates. The expressions are reported to leading order in the inter-module error probabilities. Here, we assume independent single- and two-qubit errors $\eps_1$ and $\eps_2$, respectively, that are uniformly distributed over the corresponding Pauli errors and take local gates and measurements to be perfect. For protocols with multiple outputs, we report the joint error.

\end{itemize}

\section{Demonstrations} \label{sec:demonstrations}
To benchmark the different protocols, we use fidelity-based performance metrics. For protocols producing a single Bell pair, we report the fidelity with respect to $\ket{\Phi^+}$. For protocols that output multiple Bell pairs, we compare their performance using both the joint error and a normalized per-pair fidelity derived from the joint fidelity, entabling a fair comparison across protocols with different numbers of outputs.
Specifically, if a protocol outputs $k$ Bell pairs with joint fidelity $F_{\mathrm{joint}}$ with respect to $\ket{\Phi^+}^{\otimes k}$, we define the \mbox{\textit{normalized fidelity}} as $F_{\mathrm{norm}}=F_{\mathrm{joint}}^{1/k}$ and $\eps_{\mathrm{norm}} = 1-F_{\mathrm{norm}}$. This corresponds to an effective per-pair fidelity and allows direct comparison with protocols producing a single Bell pair.

\subsection{Simulations}
\begin{figure*}[!htb]
\centering
\includegraphics[width=2\columnwidth]{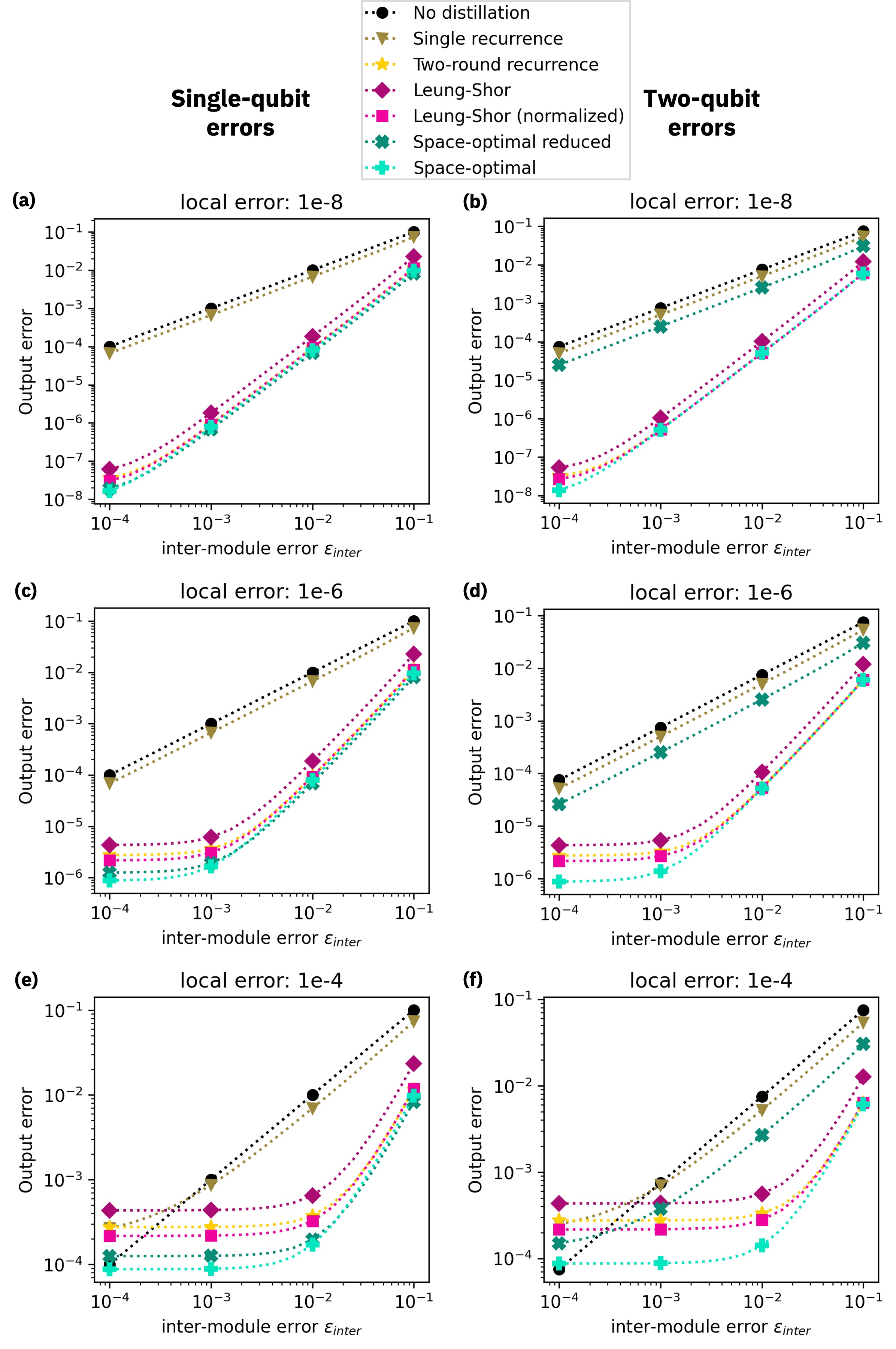}
\caption{Simulations comparing the output error of the different protocols for varying inter-module error $\eps_{inter}$. The depolarizing error of the local operations increases from $10^{-8}$ in \textbf{(a)} and \textbf{(b)} to $10^{-6}$ in \textbf{(c)} and \textbf{(d)} and to $10^{-4}$ in \textbf{(e)} and \textbf{(f)}. In \textbf{(a)}, \textbf{(c)} and \textbf{(e)} we consider only single-qubit depolarizing errors, while in \textbf{(b)}, \textbf{(d)} and \textbf{(f)} we consider uniform two-qubit depolarizing errors.} 
\label{fig:simulations}
\vspace{2cm}
\end{figure*}
To demonstrate the high potential of the entanglement distillation protocols that become more powerful for an increasing separation between inter-module and local error rates, we simulate their output errors for varying inter-module and local noise strengths in~\cref{fig:simulations}. For the local two-qubit gates, we use uniform depolarizing noise with $\eps_P=10^{-8}$ in~\cref{fig:simulations}(a)-(b), $\eps_P=10^{-6}$ in~\cref{fig:simulations}(c)-(d) and $\eps_P=10^{-4}$ in~\cref{fig:simulations}(e)-(f). For the inter-module gates, we compare two noise models: single-qubit depolarizing noise with $\eps_1=\eps_{\mathrm{inter}}$ and $\eps_2=0$ in~\cref{fig:simulations}(a),(c),(e), and uniform two-qubit depolarizing noise with $\eps_1/{6}={\eps_2}/{9}$ and $\eps_1+\eps_2=\eps_{\mathrm{inter}}$ in~\cref{fig:simulations}(b),(d),(f). 

We can clearly see the difference between the linearly scaling protocols and those with quadratic error suppression, which are converging to the lower bounds given by the local errors. Notably, across all simulations, our general space-optimal protocol shows the best performance. The reduced protocol shows the expected quadratic suppression in $\eps_1$, but only linear suppression in $\eps_2$, which emphasizes its limited utility that is confined to single-qubit errors. All other protocols show a fairly invariant behavior under the two different noise models.

\subsection{Experiments on near-term devices}
\begin{figure*}[!htb]
\centering
\includegraphics[width=2\columnwidth]{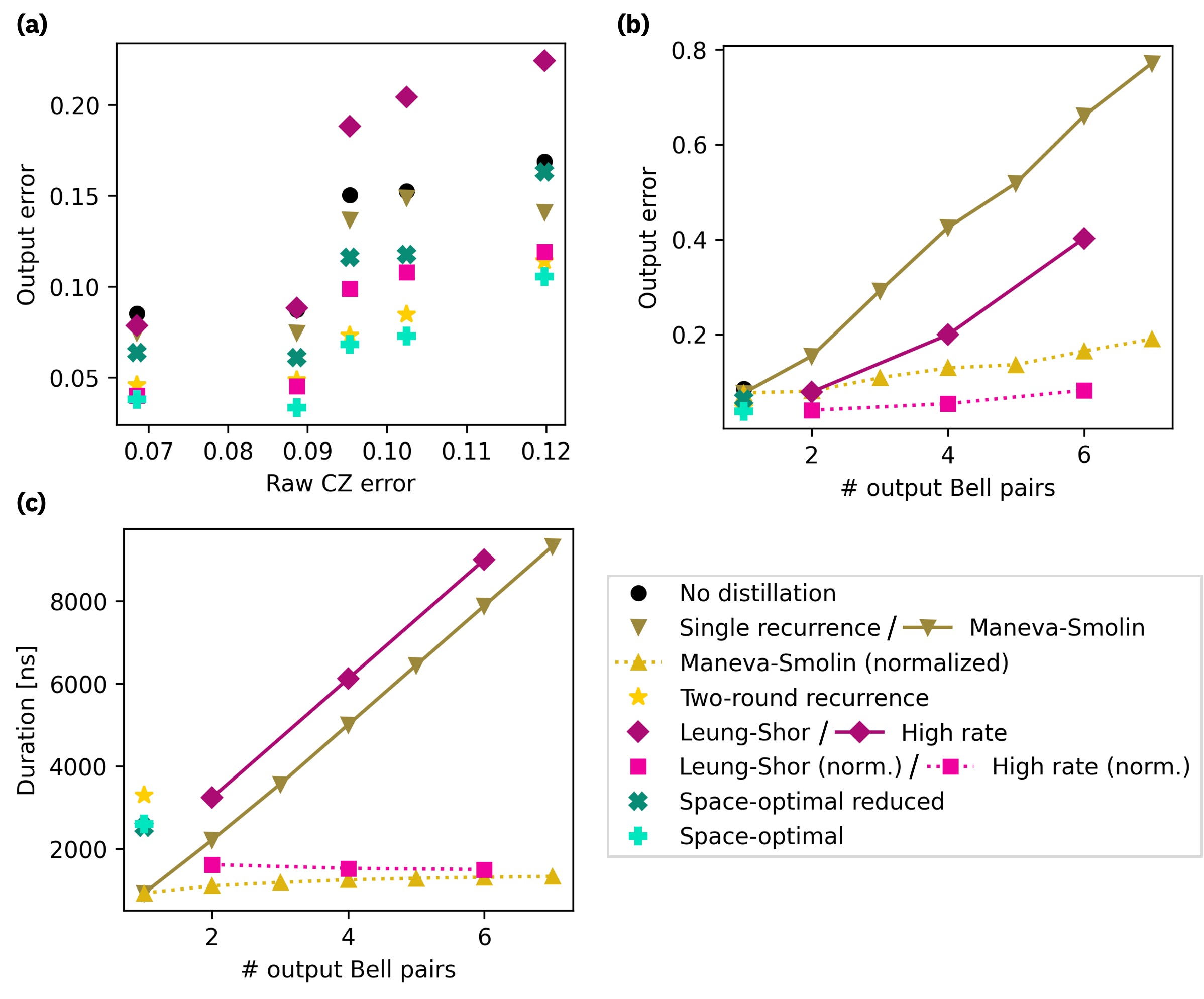}
\caption{Experimental results comparing the different protocols. \textbf{(a)} The output error for varying raw CZ errors. \textbf{(b)} The total and normalized output error for the multi-output protocols as a function of the number of Bell pairs they produce. \textbf{(c)} The total and normalized duration of the protocols when compiled on \texttt{ibm\_pittsburgh} as a function of the number of Bell pairs that are produced. Note that the single-recurrence protocol corresponds to the Maneva-Smolin protocol with one output Bell pair and similarly the Leung-Shor protocol corresponds to the high-rate protocol with two output Bell pairs.} 
\label{fig:expresults}
\end{figure*}
To demonstrate the feasibility and compare the practical utility of the protocols, we benchmark them on particularly noisy links of \texttt{ibm\_pittsburgh}. This extends recent experimental studies of entanglement distillation under realistic device noise~\cite{Siddhu2025} by including protocols with quadratic error suppression.
We take the reported CZ error of the corresponding links as \textit{raw CZ error} and compare the resulting output error on the final Bell pair when using no distillation, single Z-recurrence, two-round recurrence, the Leung-Shor protocol and our space-optimal protocol and its reduced version. For the protocols where space-time tradeoffs are possible, we have implemented the space-efficient version. The experimental results are shown in~\cref{fig:expresults}(a). The implementation details can be found in~\cref{sec:expdetails}. 

Across all tested links, the best-performing protocol is our space-optimal protocol, as it consistently yields the lowest output error. The reduced version performs worse, indicating that the errors on the CZ gates are not dominated solely by single-qubit errors components $\eps_1$, but include a non-negligible two-qubit component $\eps_2$. We can also observe that single recurrence gives only very small improvement over no distillation. This is to be expected, as it can only detect certain errors and thus even with noise-free local operations would yield an improvement only from $\eps_B$ to $\frac{2}{3}\eps_B$. However, when adding the second recurrence round, we can see that the error of the two-round recurrence improves substantially, as expected due to the quadratic error suppression. A similar performance can be observed from the normalized Leung-Shor protocol. Since the Leung-Shor protocol produces two Bell pairs, we compare it using the normalized fidelity defined above. 
The fluctuating performance compared to the two-round recurrence could be explained by the varying quality of the two additional qubits that are needed for this protocol. Note that the output error rate is always lower bounded by the combined error of the local operations, which in these experiments is only roughly one order of magnitude better.

In~\cref{fig:expresults}(b) we compare the total and normalized output error of the final Bell pairs for multi-output protocols for an increasing number of Bell pairs. While the total error is increasing as expected due to the increasing duration of the protocol, it may seem counterintuitive that also the normalized error increases for a larger number of output Bell pairs. The main reason is the linear connectivity used here that increases the duration drastically, making smaller protocols more efficient. For an increased connectivity we expect higher-output protocols to become more performant. We can observe the clearly improved fidelity for our high-rate protocol compared to the Maneva-Smolin protocol though, again as expected due to the quadratic error suppression. Note that the single recurrence protocol corresponds to the Maneva-Smolin protocol with one output Bell pair and the Leung-Shor protocol corresponds to the high-rate protocol with two output Bell pairs.

In~\cref{fig:expresults}(c) we compare the total and normalized duration of all protocols when compiled to a line on \texttt{ibm\_pittsburgh}. Note that as described in~\cref{sec:comparison}, we account only for the time until we can use the Bell pair(s). While the Maneva-Smolin protocol is the fastest, its normalized duration converges to that of our high-rate protocol. Among the protocols with quadratic error suppression, our space-optimal protocols achieve the shortest overall runtime, while the Leung-Shor protocol yields a lower time per output Bell pair.

\section{Conclusion} \label{sec:conclusion}
We presented efficient entanglement distillation protocols and evaluated their performance both theoretically and experimentally. In contrast to previous experimental studies, which focused on protocols with linear error suppression~\cite{Siddhu2025}, we investigated protocols with quadratic error suppression, leading to substantial improvements in the output fidelities. 

Our main contribution is a space-optimal protocol that requires only two qubits per module while still achieving quadratic suppression of the input noise. It is also among the fastest known distillation protocols with quadratic error suppression. Across simulations with varying inter-module and local error rates, as well as experiments on near-term devices, this protocol consistently achieves the best performance among the protocols benchmarked here. These results highlight the potential of small-scale, inter-module-gate-assisted entanglement distillation as a practical approach for overcoming noisy interconnections in modular near-term and early fault-tolerant quantum computing architectures.

We further introduced a high-rate protocol that, given $m$ qubits per module, creates $m-2$ Bell pairs with quadratically suppressed error. This protocol achieves a lower time per output Bell pair than the single-output protocols and outperforms the multi-output protocol by Maneva and Smolin~\cite{Maneva2000} in terms of error suppression.

While we demonstrated these protocols experimentally for bridging particularly noisy links within a chip, their advantage is expected to increase as the gap between local and inter-module errors grows. In architectures where inter-module connections between chips or dilution refrigerators are orders of magnitude noisier than local operations, these protocols provide a promising route toward improving interconnect fidelities. 
We note, however, that at the logical level their applicability depends on the locality of errors induced by the inter-module operations. In encodings such as the Gross code~\cite{Yoder25tourdegross}, where logical operators have extensive support over many of the 144 physical qubits, a single inter-module measurement error can induce correlated errors across multiple logical qubits and thus cannot be effectively suppressed by distillation. 
When applied at the physical level, however, entanglement distillation remains a broadly applicable and powerful technique for improving inter-module operations across quantum computing architectures.

\acknowledgments
We thank Patrick Rall, Stefan Woerner, Kevin Smith, Christopher Kang, Vikesh Siddhu and John Smolin for valuable discussions and feedback. We thank Andrew Cross for an error analysis of the Gross code. We also thank Gavin Jones and Maika Takita for their support with implementing the experiments.

%\bibliography{bibliography}
%

\appendix

\section{Experimental details} \label{sec:expdetails}
\begin{figure*}[!htb]
\centering
\includegraphics[width=2\columnwidth]{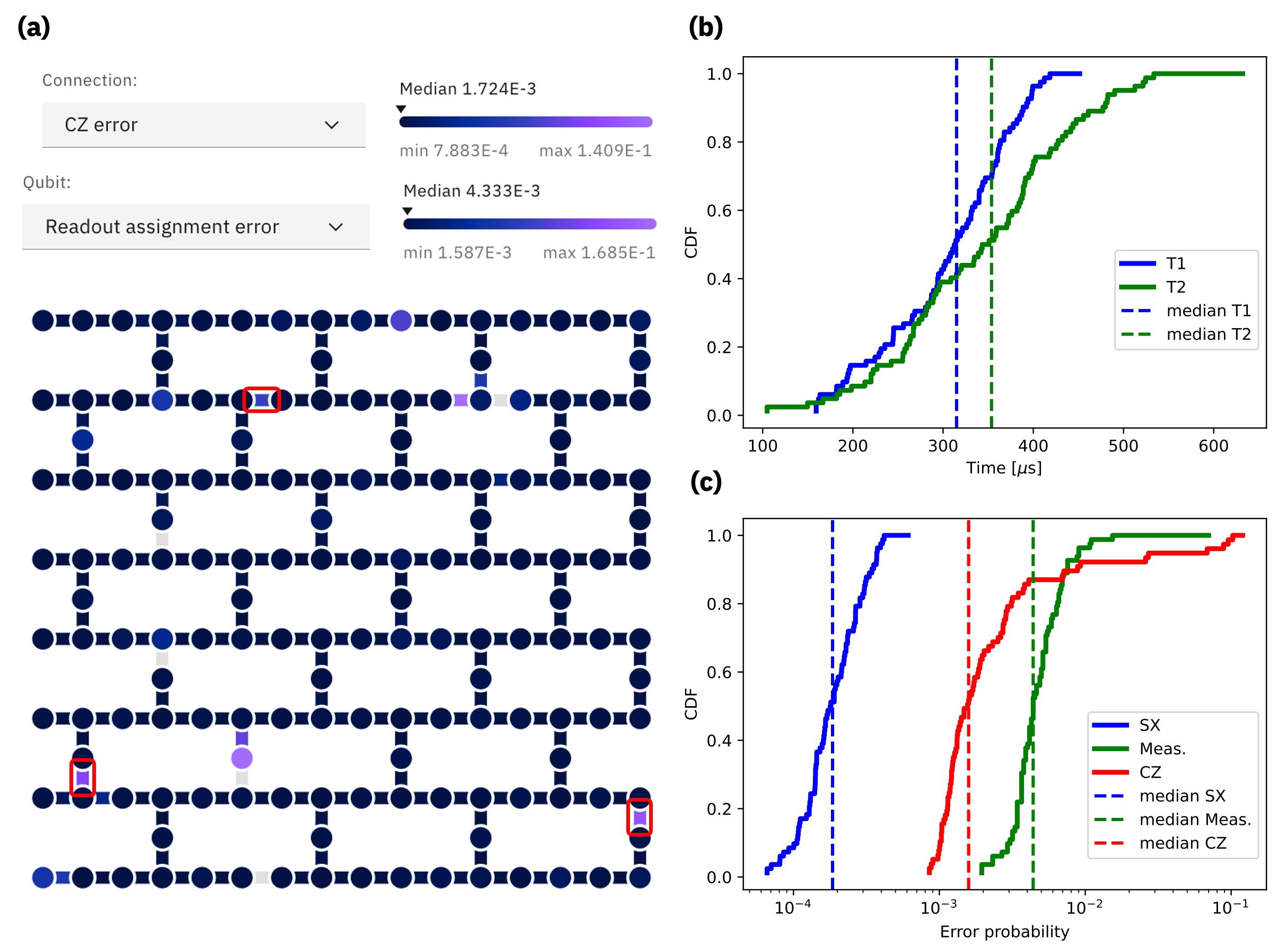}
\caption{Implementation details of the experiments. \textbf{(a)} The device layout of \texttt{ibm\_pittsburgh}, with the three noisy links used in the experiments indicated by a red box. \textbf{(b)} The cumulative distribution of the T1 and T2 coherence times and \textbf{(c)} of the single qubit gate (SX), readout (Meas.) and two-qubit gate (CZ) error rates of the qubits used in the experiments, as well as the corresponding median values.} 
\label{fig:expdetails}
\vspace{2cm}
\end{figure*}
We perform all experiments on \texttt{ibm\_pittsburgh}, a 156-qubit superconducting IBM Quantum Heron processor. 
The three noisy links chosen for the entanglement distillation experiments are indicated by a red box in~\cref{fig:expdetails}(a). Note that over time the noise changed, leading to the five experiments with different raw CZ errors in~\cref{fig:expresults}(a). The cumulative distribution of the T1 and T2 coherence times, as well as of the different error rates of all qubits used in any of the experiments are shown in~\cref{fig:expdetails}(b)-(c), indicating also the corresponding median values. 
The two-qubit gate length is $88$~ns, the single-qubit gate length is $32$~ns, the final readout length is $2584$~ns and the fast M2 readout length used for mid-circuit measurements is $1288$~ns. The fidelity of the output Bell states was estimated via direct measurements of the Pauli observables $XX$, $YY$ and $ZZ$. From these expectation values, the Bell-state fidelity with respect to $\ket{\Phi^+}$ was reconstructed. No error mitigation techniques were applied. For each circuit $10{,}000$ measurement shots were used.
\end{document}